# Les index pour les entrepôts de données : comparaison entre index arbre-B et Bitmap

El Amin Aoulad Abdelouarit  - Faculté des sciences de Tétouan – elamin@hotmail.es

Mohamed Merouani – Faculté poly disciplinaire de Tétouan – m_merouani@yahoo.fr

*Résumé* — Avec le développement des systèmes de décisionnel en générale et les entrepôts de données de manière particulière, il est devenu primordiale d'avoir une visibilité de la conception de l'entrepôt de données avant sa création, et cela vu l'importance de l'entrepôt de données qui se considère la source unique des données donnant sens à la décision. Dans un système de décisionnel, le bon fonctionnement d'un entrepôt de données réside dans le bon déroulement de l'étape intergicielle des outils ETC d'une part, et l'étape de restitution via le Datamining, des progiciels de reporting, des tableaux de bords…etc d'autre part. La grande volumétrie des données qui passes par ces étapes exigent une conception optimale pour un système de décisionnel très performant, sans négliger le choix des technologies y introduit pour la mise en place d'un entrepôt de données tel que : le système de gestion de base de données, la nature des systèmes d'exploitations du serveur, l'architecture physique des serveurs (64 bits par exemple) qui peuvent être un avantage de performance du dit système. Le concepteur de l'entrepôt de données devra prendre en considération l'efficacité de l'interrogation des données, cette efficacité repose sur la sélection d'index pertinents et leur combinaison avec les vues matérialisé, à noter que La sélection d'index est un problème NP-complet car le nombre d'index est exponentiel en nombre total d'attributs dans la base, il faut donc prévoir, lors de la conception d'un entrepôt de données, le type d'index convenable à cet entrepôt.

Cet article présente une étude comparative sur entre les index de type arbre B et de type Bitmap, leurs avantages et inconvénients, avec une expérimentation réelle démontrant que les index de type Bitmap son plus avantageux que les index de type arbre-B.

**Mots clés : Entrepôt de données, SGBD, index, business intelligence.**

*Abstract*— With the development of decision systems and specially data warehouses, the visibility of the data warehouse design before its creation has become essential, and that because of data warehouse importance as considered as the unique data source giving meaning to the decision. In a decision system the proper functioning of a data warehouse resides in the smooth running of the middleware tools ETC step one hand, and the restitution step through the data mining, reporting solutions, dashboards… etc other. The large volume of data that passes through these stages require an optimal design for a highly efficient decision system, without disregarding the choice of technologies that are introduced for the data warehouse implementation such as: database management system, the type of server operating systems, physical server architecture (64-bit, for example) that can be a benefit performance of this system. The designer of the data warehouse should consider the effectiveness of data query, this depends on the selection of relevant indexes and their combination with the materialized views, note that the index selection is a NP-complete problem, because the number of indexes is exponential in the total number of attributes in the database, So, it is necessary to provide, while the data warehouse design, the suitable type of index for this data warehouse.

This paper presents a comparative study between the index B-tree type and type Bitmap, their advantages and disadvantages, with a real experiment showing that its index of type Bitmap more advantageous than the index B-tree type.

**Keywords: Data Warehouse DBMS, indexes, business intelligence.**

# I- Introduction

L'administrateur de l'entrepôt de données prend plusieurs décisions concernant ses tâches d'administration, telle que la conception logique ou physique des bases de données, la gestion de l'espace de stockage et le réglage des performances (*performance tuning*). La plus importantes tâche est la conception physique des bases de données, qui inclut l'organisation des données et l'amélioration de l'accès à ces données. Pour améliorer les temps d'accès, l'administrateur emploie en général des index pour rechercher rapidement les informations nécessaires à une requête sans parcourir toutes les données [1],[3],[7]. La sélection d'index est difficile car leur nombre est exponentiel en nombre total d'attributs dans la base. Donc l'index joue un rôle important dans la performance des bases de données, c'est pourquoi nous nous intéressons à cet élément de l'entrepôt de donnée, qui ce considère le centre d'intérêt du concepteur lors de l'édition et l'optimisation des requêtes de sélection. L'objectif est de minimiser le temps d'exécution des requêtes, et tant que les requêtes dans un entrepôt de données se basent sur les index, nous allons travailler sur le problème du choix de type d'index lors de la conception de notre entrepôt de données. Il existe plusieurs types pris en charge par les bases de données tel que Bitmap [4] , Arbre-B [3], [6], [7], [9], Bitmap de jointure[10], range-based bitmap index [11]etc. Dans ce sens nous avons choisis deux types d'index par pertinence pour cette étude, l'index de type : Arbre-B et index de type Bitmap.

Cet article est répartit comme suit, après la présentation du contexte et les travaux liés avec (section 2), puis une présentation des hypothèses (section 3) une expérimentation sera réalisé pour déterminer des résultats concrets (section 4). Nous terminons par une conclusion et des perspectives de recherches (section 5)

## II- Présentation de l'axe de recherche

### 1. Index Bitmap

#### 1.1 Définition et exemple:

Un index bitmap est une structure de données définie dans un SGBD, utilisée pour optimiser l'accès aux données dans les bases de données. C'est un type d'indexation qui est particulièrement intéressant et performant dans le cadre des requêtes de sélection. L'index bitmap d'un attribut est codé sur des bits, d'où son faible coût en terme d'espace occupé. [7] Toutes les valeurs possibles de l'attribut sont considérées, que la valeur soit présente ou non dans la table. A chacune de ces valeurs correspond un tableau de bits, appelé bitmap, qui contient autant de bits que de n-uplets présents dans la table. Ainsi, ce type d'index est très efficace lorsque les attributs ont un faible nombre de valeurs distinctes. Chaque bit représente donc la valeur d'un attribut, pour un n-uplet donné. Pour chacun des bits, il y a un codage de présence/absence (1/0), ce qui traduit le fait qu'un n-uplet présente ou non la valeur caractérisée par le bitmap.

| ROWID | C | B0 | B1 | B2 | B3 |
|-------|---|----|----|----|----|
| 0 | 2 | 0 | 0 | 1 | 0 |
| 1 | 1 | 0 | 1 | 0 | 0 |
| 2 | 3 | 0 | 0 | 0 | 1 |
| 3 | 0 | 1 | 0 | 0 | 0 |
| 4 | 3 | 0 | 0 | 0 | 1 |
| 5 | 1 | 0 | 1 | 0 | 0 |
| 6 | 0 | 1 | 0 | 0 | 0 |
| 7 | 0 | 1 | 0 | 0 | 0 |
| 8 | 2 | 0 | 0 | 1 | 0 |

Tableau 1 Bitmap basic adopté par [10]

Pour illustrer le fonctionnement d'un index Bitmap on prend un exemple E.E-O'Neil et P.P-O'Neil [12]. « Table 1 illustre un index de type bitmap basic dans une table contenant 9 enregistrements, où l'index est créé dans la colonne C avec des entiers allant de 0 à 3, on dit que la cardinalité de la colonne C est 4, par ce que on a 4 valeurs distincts [0, 1, 2, 3], d'où l'index bitmap de C contiens 4 Bitmaps indiqués comme B0, B1, B2 et B3 correspondant la valeur represented. Dans cet exemple, la première ligne où RowID=0, la colonne C est de valeur 2, par conséquence, la colonne B2 est de valeur de bit « 1 », pendant que les autres bitmaps sont mis à « 0 ». Même chose pour la ligne suivante, où C=1 qui correspond à que le bitmap B1 est mis sur 1 et le reste à « 0 ». Ce processus se répète pour le reste des lignes. [12]

### 1.2 Propriétés :

Les index bitmap possèdent une propriété très intéressante qui consiste à répondre à certains types de requêtes sans retourner aux données elles-mêmes, optimisant ainsi les temps de réponse. Cela est possible grâce aux opérations de comptage (COUNT) et aux opérateurs logiques (AND, OR, etc) qui agissent "bit à bit" sur les bitmaps.

## 2. Index B-Arbre

### 2.1 Définition

L'index B-arbre stocke les pointeurs d'index et les valeurs à d'autres nœuds d'index en utilisant une structure d'arbre récursive. [3], [6], [7], [9], Les données sont facilement relevées par les traces des pointeurs. Le plus haut niveau de l'index est appelé racine pendant que le niveau le plus bas on l'appelle nœud de feuille ou « leaf node ». [7] Tous les autres niveaux entre eux sont appelés des branches ou nœuds internes. Toutes les racines et les branches contiennent des entrées qui pointent vers le niveau suivant de l'index. Nœuds feuilles se composent de la clé d'index et des pointeurs pointant vers l'emplacement physique des enregistrements. Nous présentons plus de détails sur la structure d'index arbre-B[7].

La structure B-Arbre est utilisée par le serveur de base de données pour configurer l'index (Figure1)

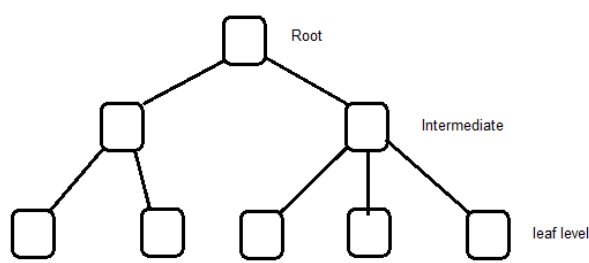

**Root ou racine:** plus haut niveau de l'index pointe vers les niveaux suivants des nodes branches.

**Nœuds intermédiaires ou branches :** contiennent des pointeurs vers le niveau suivant des branches ou vers les nœuds de feuilles.

**Nœud de feuilles ou leaf nodes :** le plus bas niveau de l'index pointe vers d'autres nœud de feuilles.

## III- Hypothèses

La sagesse conventionnelle veut que les index bitmap soient les plus appropriés pour les colonnes ayant des valeurs distinctes faibles - comme le sexe, état civil, et la relation. Cette hypothèse n'est pas tout à fait exacte, cependant. En réalité, un index bitmap est toujours conseillé pour les systèmes dans lesquels les données ne sont pas souvent mises à jour par de nombreux systèmes concurrents. En fait, comme je vais démontrer ici, un index bitmap sur une colonne avec des valeurs uniques à 100% (un candidat de la colonne de clé primaire) est aussi efficace qu'un index de B-arbre.

Dans cet article je vais vous donner quelques exemples, ainsi que les décisions de l'optimiseur, qui sont communs aux deux types d'index sur une colonne de faible cardinalité ainsi qu'une grande cardinalité un. Ces exemples aideront les DBA à comprendre que l'utilisation des index bitmap n'est pas en fait cardinal dépendante mais dépend de l'application.

## IV- Expérimentation :

L'expérimentation se déroule dans un serveur de base de données Oracle supportant OLAP. Il y en a plusieurs inconvénients en utilisant un index de type bitmap dans une colonne unique, l'un est le besoin de plus d'espace de disque. Cependant, la taille de l'index bitmap dépend de la cardinalité de la colonne sur laquelle il est créé, ainsi que la répartition des données. Par conséquent, un index bitmap sur la colonne « Genre » sera inférieur à un index B-arbre sur la même colonne. En revanche, un index bitmap sur EMPNO (candidat à être clé primaire) sera beaucoup plus grand qu'un index B-arbresur cette colonne. Mais parce que moins d'utilisateurs accèdent aux systèmes décisionnel (DSS) qu'en accédant les systèmes de traitement transactionnel (OLTP), les ressources ne posent pas de problème pour ces applications.

Pour illustrer ce point, j'ai créé deux tables, TEST_NORMAL et TEST_RANDOM. On a inséré un million de lignes dans la table TEST_NORMAL l'aide d'un bloc PL / SQL, puis on a inséré ces lignes dans la table de TEST_RANDOM dans un ordre aléatoire: [Tableau2]

Notez que la table TEST_NORMAL est organisé et que la table TEST_RANDOM créée au hasard, elle a donc des données désorganisées. Dans la table ci-dessus, la colonne EMPNO a des valeurs distinctes à 100%, elle est donc un bon candidat pour devenir une clé primaire. Si vous définissez cette colonne comme clé primaire, vous allez créer un index B-arbre et non pas un index bitmap parce que Oracle ne supporte pas les index bitmap de clé primaire.

```
Create table test_normal (empno number(10),
ename varchar2(30), sal number(10));

Begin
For i in 1..1000000
Loop
   Insert into test_normal
   values(i,    dbms_random.string('U',30),
dbms_random.value(1000,7000));
   If mod(i, 10000) = 0 then
   Commit;
  End if;
End loop;
End;
/

Create table test random
as
select /*+ append */ * from test_normal
order by dbms_random.random;

SQL> select count(*) "Total  Rows"  from
test_normal;

Total Rows
----------
    1000000

Elapsed: 00:00:01.09

SQL> select count(distinct empno) "Distinct
Values" from test_normal;

Distinct Values
---------------
        1000000

Elapsed: 00:00:06.09
SQL> select  count(*)  "Total  Rows"  from
test random;

Total Rows
----------
    1000000

Elapsed: 00:00:03.05
SQL> select count(distinct empno) "Distinct
Values" from test random;

Distinct Values
---------------
        1000000

Elapsed: 00:00:12.07
```

Tableau 2

Pour analyser le comportement de ces index, nous allons effectuer les étapes suivantes:

1. Sur TEST_NORMAL :

   a. Créer un index bitmap sur la colonne EMPNO et exécuter des requêtes avec prédicats d'égalité.

   b. Créer un index B-Arbre sur la colonne EMPNO, exécuter des requêtes avec prédicats d'égalité, et de comparer le physique et logique E/S fait par les requêtes pour récupérer les résultats pour les différents ensembles de valeurs.

   2. Sur TEST_RANDOM:

   a. Identique à l'étape 1A.

   b. Identique à l'étape 1b.

3. Sur TEST_NORMAL:

   a. Identique à l'étape 1A, sauf que les requêtes sont exécutées dans un intervalle de prédicats.

   b. Identique à l'étape 1b, sauf que les requêtes sont exécutées dans une fourchette de prédicats. Et après on compare les statistiques.

4. Sur TEST_RANDOM:

   a. Même que l'étape 3A.

   b. Identique à l'étape 3B.

5. Sur TEST_NORMAL

   a. Créer un index bitmap sur la colonne de salaire SAL, puis exécuter des requêtes avec prédicats d'égalité et certaines avec des intervalles de prédicats.

   b. Créer un index B-Arbre sur la colonne SAL, puis exécuter des requêtes avec prédicats d'égalité et certaines avec des intervalles de prédicats (même ensemble de valeurs que dans l'étape 5A). Comparer les E/S fait par les requêtes pour récupérer les résultats.

6. Ajouter une colonne « genre » pour les deux tables, et mettre à jour la colonne avec trois valeurs possibles: M pour masculin, F pour féminin, et nulle pour N/A. Cette colonne est mise à jour avec ces valeurs en fonction de certaines conditions.

7. Créer un index bitmap sur cette colonne, puis exécuter des requêtes avec prédicats d'égalité.

8. Créer un index B-Arbre sur la colonne de genre, puis exécuter des requêtes avec prédicats d'égalité. Comparer les résultats de l'étape 7.

Les étapes 1 à 4 impliquent un haut cardinal (distincts à 100%) colonne, étape 5, une colonne normale cardinal, et les étapes 7 et 8 une colonne basse cardinal.

**Étape 1A (sur TEST_NORMAL) :**

Dans cette étape, nous allons créer un index bitmap sur la table TEST_NORMAL puis vérifiez la taille de cet index, son coefficient de regroupement, et la taille de la table. Ensuite, nous allons exécuter des requêtes avec prédicats d'égalité et de noter les

entrées / sorties de ces requêtes en utilisant cet index bitmap. (Table 3)

```
SQL> create bitmap index normal_empno_bmx
on test_normal(empno);

Index created.

Elapsed: 00:00:29.06
SQL> analyze table test_normal compute
statistics for table for all indexes for
all indexed columns;

Table analyzed.

Elapsed: 00:00:19.01
SQL>    select    substr(segment_name,1,30)
segment_name, bytes/1024/1024 "Size in MB"
  2 from user_segments
  3*      where      segment_name      in
('TEST_NORMAL','NORMAL_EMPNO_BMX');

SEGMENT NAME            Size in MB
------------------------------------
---------------
TEST_NORMAL             50
NORMAL_EMPNO_BMX        28

Elapsed: 00:00:02.00
SQL> select index_name, clustering_factor
from user_indexes;

INDEX NAME              CLUSTERING FACTOR
------------------------------    ----
------------------------------
NORMAL_EMPNO_BMX        1000000

Elapsed: 00:00:00.00
```

Tableau 3

Vous pouvez le voir dans le tableau ci-dessus que la taille de l'index est 28 Mo et que le facteur de regroupement est égal au nombre de lignes dans le tableau. Maintenant, nous allons exécuter les requêtes avec prédicats d'égalité pour les différents ensembles de valeurs: (Tableau 4)

**Etape 1B (sur TEST_NORMAL)**

Maintenant, nous allons laisser tomber cet index bitmap et créer un index B-arbresur la colonne EMPNO. Comme précédemment, nous allons vérifier la taille de l'index et de son facteur de regroupement et d'exécuter les mêmes requêtes pour le même ensemble de valeurs, de comparer les E/S.

```
SQL> set autotrace only
SQL> select * from test_normal where
empno=&empno;
Enter value for empno: 1000
old   1: select * from test_normal where
empno=&empno
new   1: select * from test_normal where
empno=1000

Elapsed: 00:00:00.01

Execution Plan
----------------------------------------------
--------------
   0      SELECT STATEMENT Optimizer=CHOOSE
(Cost=4 Card=1 Bytes=34)
   1    0    TABLE ACCESS (BY INDEX ROWID) OF
'TEST_NORMAL' (Cost=4 Car
        d=1 Bytes=34)
   2    1    BITMAP CONVERSION (TO ROWIDS)
   3    2        BITMAP INDEX (SINGLE VALUE)
OF 'NORMAL_EMPNO_BMX'

Statistics
----------------------------------------------
--------------
        0  recursive calls
        0  db block gets
        5  consistent gets
        0  physical reads
        0  redo size
      515  bytes sent via SQL*Net to
client
      499  bytes received via SQL*Net from
client
        2  SQL*Net roundtrips to/from
client
        0  sorts (memory)
        0  sorts (disk)
        1  rows processed
```

Tableau 4

```
SQL> drop index NORMAL_EMPNO_BMX;

Index dropped.

SQL> create index normal_empno_idx on
test_normal(empno);

Index created.

SQL> analyze table test_normal compute
statistics for table for all indexes for
all indexed columns;

Table analyzed.

SQL>    select    substr(segment_name,1,30)
segment_name, bytes/1024/1024 "Size in MB"
  2 from user_segments
  3        where      segment_name      in
('TEST_NORMAL','NORMAL_EMPNO_IDX');

SEGMENT_NAME            Size in MB
----------------------------------
TEST_NORMAL             50
NORMAL_EMPNO_IDX        18

SQL> select index_name, clustering_factor
from user_indexes;

INDEX NAME              CLUSTERING FACTOR
----------------------------------    -----
------------------------------
NORMAL_EMPNO_IDX        6210
```

Tableau 5

Il est clair dans ce tableau (Tableau 5) que les index B-Arbres est inférieur que l'index bitmap sur la colonne EMPNO. Le facteur de regroupement de l'index B-arbre est beaucoup plus proche du nombre de blocs dans une table, pour cette raison, l'index B-arbre est efficace pour les requêtes de prédicats. Maintenant, nous allons exécuter les mêmes requêtes pour le même ensemble de valeurs, en utilisant notre index B-arbre.

```
SQL> set autotrace
SQL>   select  *  from  test_normal  where
empno=&empno;
Enter value for empno: 1000
old   1: select * from test normal where
empno=&empno
new   1: select * from test_normal where
empno=1000

Elapsed: 00:00:00.01

Execution Plan
---------------------------------------------
   0      SELECT STATEMENT Optimizer=CHOOSE
(Cost=4 Card=1 Bytes=34)
   1    0    TABLE ACCESS (BY INDEX ROWID) OF
'TEST_NORMAL' (Cost=4 Car
        d=1 Bytes=34)
   2    1          INDEX (RANGE SCAN) OF
'NORMAL EMPNO IDX' (NON-UNIQUE) (C
            ost=3 Card=1)

Statistics
---------------------------------------------
29  recursive calls
        0  db block gets
        5  consistent gets
        0  physical reads
        0  redo size
      515    bytes sent  via  SQL*Net  to
client
      499  bytes received via SQL*Net from
client
        2    SQL*Net  roundtrips  to/from
client
        0  sorts (memory)
        0  sorts (disk)
        1  rows processed
```
Tableau 6

Comme vous pouvez le voir (Tableau 7), lorsque les requêtes sont exécutées pour des valeurs différentes, le nombre de lectures cohérentes et des lectures physiques sont identiques pour bitmap et index B-arbre sur une colonne unique de 100%.

| BITMAP | | EMPNO | B-TREE | |
|---|---|---|---|---|
| Consistent Reads | Physical Reads | | Consistent Reads | Physical Reads |
| 5 | 0 | 1000 | 5 | 0 |
| 5 | 2 | 2398 | 5 | 2 |
| 5 | 2 | 8545 | 5 | 2 |
| 5 | 2 | 98008 | 5 | 2 |
| 5 | 2 | 85342 | 5 | 2 |
| 5 | 2 | 128444 | 5 | 2 |
| 5 | 2 | 858 | 5 | 2 |

Tableau 7

## Etape 2A (sur TEST_RANDOM)

Maintenant, nous allons effectuer la même expérience sur TEST_RANDOM (Tableau 8):

```
SQL> create bitmap index random_empno_bmx
on test_random(empno);

Index created.

SQL>  analyze  table  test_random  compute
statistics  for  table  for  all  indexes  for
all indexed columns;

Table analyzed.

SQL>    select    substr(segment_name,1,30)
segment_name, bytes/1024/1024 "Size in MB"
  2  from user_segments
  3*      where      segment name      in
('TEST RANDOM','RANDOM EMPNO BMX');

SEGMENT_NAME               Size in MB
------------------------------------
---------------
TEST RANDOM                  50
RANDOM EMPNO BMX             28

SQL>  select  index_name,  clustering_factor
from user_indexes;

INDEX NAME               CLUSTERING FACTOR
------------------------------   ----
----------------------------
RANDOM_EMPNO_BMX             1000000
```
Tableau 8

Encore une fois, les statistiques (taille et le facteur de regroupement) sont identiques à celles de l'index sur la table TEST_NORMAL (Tableau 9):

```
Elapsed: 00:00:00.01

Execution Plan
---------------------------------------------
---------------
   0      SELECT STATEMENT Optimizer=CHOOSE
(Cost=4 Card=1 Bytes=34)
   1    0    TABLE ACCESS (BY INDEX ROWID) OF
'TEST_RANDOM' (Cost=4 Card=1 Bytes=34)
   2    1    BITMAP CONVERSION (TO ROWIDS)
   3    2        BITMAP INDEX (SINGLE VALUE)
OF 'RANDOM EMPNO BMX'

Statistics
---------------------------------------------
---------------
        0  recursive calls
        0  db block gets
        5  consistent gets
        0  physical reads
        0  redo size
      515    bytes sent  via  SQL*Net  to
client
      499  bytes received via SQL*Net from
client
        2    SQL*Net  roundtrips  to/from
client
        0  sorts (memory)
        0  sorts (disk)
        1  rows processed
```
Tableau 9

## Étape 2B (sur TEST_RANDOM)

Maintenant, comme dans l'étape 1b, nous allons supprimer l'index bitmap et créer un index B-arbre sur la colonne EMPNO. (Tableau 10)

```
SQL> drop index RANDOM EMPNO BMX;

Index dropped.

SQL>  create  index  random_empno_idx  on
test_random(empno);

Index created.

SQL>  analyze  table  test_random  compute
statistics  for  table  for  all  indexes  for
all indexed columns;

Table analyzed.

SQL>    select    substr(segment_name,1,30)
segment_name, bytes/1024/1024 "Size in MB"
  2  from user_segments
  3        where    segment name    in
('TEST RANDOM','RANDOM EMPNO IDX');

SEGMENT_NAME           Size in MB
------------------------------
----------------
TEST RANDOM            50
RANDOM_EMPNO_IDX       18

SQL> select index_name, clustering_factor
from user_indexes;

INDEX NAME           CLUSTERING FACTOR
------------------------------  -----
------------------------------
RANDOM EMPNO IDX     999830
```
Tableau 10

Ce tableau montre que la taille de l'index est égale à la taille de cet index sur la table TEST_NORMAL mais le facteur de regroupement est beaucoup plus proche du nombre de lignes, ce qui rend cet index inefficace pour les requêtes de prédicat par intervalles (que nous verrons à l'étape 4).

Ce facteur de regroupement n'aura aucune influence sur l'égalité des requêtes de prédicat parce que les lignes ont des valeurs 100% distinctes et le nombre de lignes par clé est 1.

Maintenant, nous allons exécuter les requêtes avec prédicats d'égalité et le même ensemble de valeurs. (Tableau11)

Encore une fois, les résultats sont presque identiques à celles dans les étapes 1A et 1B. La distribution de données n'affecte pas le montant des lectures cohérentes et des lectures physiques d'une colonne unique.

```
SQL>  select  *  from  test_random  where
empno=&empno;
Enter value for empno: 1000
old   1: select * from test random where
empno=&empno
new   1: select * from test random where
empno=1000

Elapsed: 00:00:00.01

Execution Plan
----------------------------------------------
--------------
   0       SELECT STATEMENT Optimizer=CHOOSE
(Cost=4 Card=1 Bytes=34)
   1   0   TABLE ACCESS (BY INDEX ROWID) OF
'TEST RANDOM' (Cost=4 Card=1 Bytes=34)
   2   1   INDEX  (RANGE  SCAN)  OF
'RANDOM_EMPNO_IDX'    (NON-UNIQUE)   (Cost=3
Card=1)

Statistics
----------------------------------------------
--------------
        0  recursive calls
        0  db block gets
        5  consistent gets
        0  physical reads
        0  redo size
      515  bytes  sent  via  SQL*Net  to
client
      499  bytes received via SQL*Net from
client
        2  SQL*Net  roundtrips  to/from
client
        0  sorts (memory)
        0  sorts (disk)
        1  rows processed
```
Tableau 11

### Etape 3A (sur TEST_NORMAL)

Dans cette étape, nous allons créer l'index bitmap (similaire à l'étape 1A). Sachant que la taille et le facteur de regroupement de l'index est égal au nombre de lignes dans la table. Maintenant, nous allons lancer quelques requêtes avec intervalles de prédicats (Tableau 12).

### Étape 3B (sur TEST_NORMAL)

Dans cette étape, nous allons exécuter les requêtes sur la table TEST_NORMAL avec un index B-arbre sur elle. (Tableau 13)

| BITMAP | | EMPNO | B-TREE | |
|---|---|---|---|---|
| Consistent Reads | Physical Reads | (Range) | Consistent Reads | Physical Reads |
| 331 | 0 | 1-2300 | 329 | 0 |
| 285 | 0 | 8-1980 | 283 | 0 |
| 346 | 19 | 1850-4250 | 344 | 16 |
| 427 | 31 | 28888-31850 | 424 | 28 |
| 371 | 27 | 82900-85478 | 367 | 23 |
| 2157 | 149 | 984888-1000000 | 2139 | 35 |

Tableau 14

```
SQL> select * from test_normal where empno
between &range1 and &range2;
Enter value for range1: 1
Enter value for range2: 2300
old   1: select * from test normal where
empno between &range1 and &range2
new   1: select * from test_normal where
empno between 1 and 2300

2300 rows selected.

Elapsed: 00:00:00.02

Execution Plan
-------------------------------------------
   0      SELECT STATEMENT Optimizer=CHOOSE
(Cost=23 Card=2299 Bytes=78166)
   1   0    TABLE ACCESS (BY INDEX ROWID) OF
'TEST_NORMAL'        (Cost=23      Card=2299
Bytes=78166)
   2   1       INDEX  (RANGE  SCAN)  OF
'NORMAL EMPNO IDX'  (NON-UNIQUE)   (Cost=8
Card=2299)

Statistics
-------------------------------------------
         0  recursive calls
         0  db block gets
       329  consistent gets
        15  physical reads
         0  redo size
    111416  bytes sent via SQL*Net client
      2182  bytes received via SQL*Net from
client
       155   SQL*Net roundtrips to/from
client
         0  sorts (memory)
         0  sorts (disk)
      2300  rows processed
```

Tableau 12

```
SQL> select * from test normal where empno
between &range1 and &range2;
Enter value for range1: 1
Enter value for range2: 2300
old   1: select * from test_normal where
empno between &range1 and &range2
new   1: select * from test_normal where
empno between 1 and 2300

2300 rows selected.
Elapsed: 00:00:00.02
Execution Plan
-------------------------------------------
   0      SELECT STATEMENT Optimizer=CHOOSE
(Cost=23 Card=2299 Bytes=78166)
   1   0    TABLE ACCESS (BY INDEX ROWID) OF
'TEST NORMAL'        (Cost=23     Card=2299
Bytes=78166)
   2   1       INDEX  (RANGE   SCAN)  OF
'NORMAL_EMPNO_IDX'   (NON-UNIQUE)  (Cost=8
Card=2299)
Statistics
-------------------------------------------
         0  recursive calls
         0  db block gets
       329  consistent gets
        15  physical reads
         0  redo size
    111416   bytes  sent  via  SQL*Net  to
client
      2182  bytes received via SQL*Net from
client
       155   SQL*Net  roundtrips  to/from
client
         0  sorts (memory)
         0  sorts (disk)
      2300  rows processed
```

Tableau 13

Lorsque ces requêtes sont exécutées pour différents intervalles, on obtient ces résultats;

Comme vous pouvez le voir, le nombre de lectures cohérentes et de lectures physiques pour les deux index sont de nouveau presque identiques. Le dernier intervalle (984888-1000000) a retourné presque 15.000 lignes, ce qui était le nombre maximal de lignes extraites pour tous les intervalles indiqués ci-dessus. Ainsi, lorsque nous avons demandé un scan de table complet, la lecture cohérente et compte lire physiques étaient 7,239 et 5,663, respectivement.

**Etape 4A (sur TEST_RANDOM)**

Dans cette étape, nous allons exécuter les requêtes avec prédicats d'intervalles sur la table de TEST_RANDOM avec index bitmap et vérifier les lectures cohérentes et les lectures physiques. Ici, vous verrez l'impact du facteur de regroupement. (Tableau 15).

```
SQL>select * from test_random where empno
between &range1 and &range2;
Enter value for range1: 1
Enter value for range2: 2300
old  1: select * from test random where
empno between &range1 and &range2
new  1: select * from test_random where
empno between 1 and 2300

2300 rows selected.

Elapsed: 00:00:08.01

Execution Plan
-------------------------------------------
---------------
   0      SELECT STATEMENT Optimizer=CHOOSE
(Cost=453 Card=2299 Bytes=78166)
   1   0    TABLE ACCESS (BY INDEX ROWID) OF
'TEST_RANDOM' (Cost=453 Card=2299
Bytes=78166)
   2   1      BITMAP CONVERSION (TO ROWIDS)
   3   2        BITMAP INDEX (RANGE SCAN) OF
'RANDOM_EMPNO_BMX'

Statistics
-------------------------------------------
---------------
         0  recursive calls
         0  db block gets
      2463  consistent gets
      1200  physical reads
         0  redo size
    111416  bytes sent via SQL*Net to
client
      2182  bytes received via SQL*Net from
client
       155  SQL*Net roundtrips to/from
client
         0  sorts (memory)
         0  sorts (disk)
      2300  rows processed
```

Tableau 15

**L'étape 4B (sur TEST_RANDOM)**

Dans cette étape, nous allons exécuter les requêtes de prédicat d'intervalle sur TEST_RANDOM avec un index B-arbre. Rappelons que le facteur de regroupement de cet index était très proche du nombre de lignes dans une table (et donc inefficace). Voici ce que l'optimiseur a à dire à ce sujet: (Tableau 16)

```
SQL> select * from test random where empno
between &range1 and &range2;
Enter value for range1: 1
Enter value for range2: 2300
old   1: select * from test_random where
empno between &range1 and &range2
new   1: select * from test random where
empno between 1 and 2300

2300 rows selected.

Elapsed: 00:00:03.04

Execution Plan
-------------------------------------------
   0      SELECT STATEMENT Optimizer=CHOOSE
(Cost=613 Card=2299 Bytes=78166)
   1    0    TABLE ACCESS (FULL) OF
'TEST RANDOM' (Cost=613 Card=2299
Bytes=78166)

Statistics
-------------------------------------------
         0   recursive calls
         0   db block gets
      6415   consistent gets
      4910   physical reads
         0   redo size
    111416   bytes sent via SQL*Net to
client
      2182   bytes received via SQL*Net
from client
       155   SQL*Net roundtrips to/from
client
         0   sorts (memory)
         0   sorts (disk)
      2300   rows processed
```
Tableau 16

L'optimiseur a opté pour une analyse complète de la table plutôt qu'utiliser l'index à cause du facteur de regroupement:

| BITMAP | | EMPNO | B-TREE | |
|--------|--------|--------|--------|--------|
| Consistent Reads | Physical Reads | (Range) | Consistent Reads | Physical Reads |
| 2463 | 1200 | 1-2300 | 6415 | 4910 |
| 2114 | 31 | 8-1980 | 6389 | 4910 |
| 2572 | 1135 | 1850-4250 | 6418 | 4909 |
| 3173 | 1620 | 28888-31850 | 6456 | 4909 |
| 2762 | 1358 | 82900-85478 | 6431 | 4909 |
| 7254 | 3329 | 984888-1000000 | 7254 | 4909 |

Tableau 17

Pour le dernier intervalle (984888-1000000) seulement, l'optimiseur a opté pour une analyse complète de la table pour l'index bitmap, alors que pour tous les intervalles, il a opté pour une analyse complète de la table pour l'index B-arbre. Cette disparité est due au facteur de regroupement: L'optimiseur ne tient pas compte de la valeur du facteur de regroupement lors de la génération des plans d'exécution en utilisant un index bitmap, alors qu'il le fait que pour un index B-arbre. Dans ce scénario, l'index de bitmap remplit plus efficacement que l'index B-arbre. Les étapes suivantes révèlent des faits intéressants sur ces index :

**Etape 5A (sur TEST_NORMAL)**

Créer un index bitmap sur la colonne SAL de la table TEST_NORMAL. Cette colonne est de cardinal normal.

```
SQL> create bitmap index normal sal bmx on
test_normal(sal);

Index created.

SQL> analyze table test normal compute
statistics for table for all indexes for
all indexed columns;

Table analyzed.
```
Tableau 18

Maintenant, nous allons obtenir la taille de l'index et le facteur de regroupement.

```
SQL>select substr(segment_name,1,30)
segment name, bytes/1024/1024 "Size in MB"
  2* from user segments
  3* where segment_name in
('TEST_NORMAL','NORMAL_SAL_BMX');

SEGMENT_NAME                 Size in MB
-----------------------------
--------------
TEST_NORMAL                  50
NORMAL_SAL_BMX               4

SQL> select index name, clustering factor
from user indexes;

INDEX_NAME               CLUSTERING_FACTOR
-----------------------------    ----
-----------------------------
NORMAL SAL BMX           6001
```
Tableau 19

Pour les requêtes, première exécution avec prédicats d'égalité (Tableau 20), Et après avec prédicats d'intervalle (Tableau 21)

Maintenant, on supprime l'index bitmap et créer un index B-arbre sur TEST_NORMAL.

```
SQL> set auto trace
SQL> select * from test_normal where
sal=&sal;
Enter value for sal: 1869
old   1: select * from test_normal where
sal=&sal
new   1: select * from test_normal where
sal=1869

164 rows selected.

Elapsed: 00:00:00.08

Execution Plan
-------------------------------------------
   0      SELECT STATEMENT Optimizer=CHOOSE
(Cost=39 Card=168 Bytes=4032)
   1   0    TABLE ACCESS (BY INDEX ROWID) OF
'TEST_NORMAL' (Cost=39 Card=168 Bytes=4032)
   2   1      BITMAP CONVERSION (TO ROWIDS)
   3   2        BITMAP INDEX (SINGLE VALUE)
OF 'NORMAL_SAL_BMX'

Statistics
-------------------------------------------
          0  recursive calls
          0  db block gets
        165  consistent gets
          0  physical reads
          0  redo size
       8461  bytes sent via SQL*Net to
client
        609  bytes received via SQL*Net from
client
         12  SQL*Net roundtrips to/from
client
          0  sorts (memory)
          0  sorts (disk)
        164  rows processed
```

Tableau 20

```
SQL> select * from test_normal where sal
between &sal1 and &sal2;
Enter value for sal1: 1500
Enter value for sal2: 2000
old   1: select * from test_normal where sal
between &sal1 and &sal2
new   1: select * from test_normal where sal
between 1500 and 2000

83743 rows selected.
Elapsed: 00:00:05.00

Execution Plan
-------------------------------------------
   0      SELECT STATEMENT Optimizer=CHOOSE
(Cost=601 Card=83376 Bytes
=2001024)
   1   0    TABLE ACCESS (FULL) OF
'TEST_NORMAL' (Cost=601 Card=83376
          Bytes=2001024)

Statistics
-------------------------------------------
          0  recursive calls
          0  db block gets
      11778  consistent gets
       5850  physical reads
          0  redo size
    4123553  bytes sent via SQL*Net client
      61901  bytes received via SQL*Net from
client
       5584  SQL*Net roundtrips to/from
client
          0  sorts (memory)
          0  sorts (disk)
      83743  rows processed
```

Tableau 21

```
SQL> create index normal_sal_idx on
test_normal(sal);

Index created.

SQL> analyze table test_normal compute
statistics for table for all indexes for
all indexed columns;

Table analyzed.
```

**Take a look at the size of the index and the clustering factor.**

```
SQL> select substr(segment_name,1,30)
segment_name, bytes/1024/1024 "Size in MB"
  2  from user_segments
  3  where segment_name in
('TEST_NORMAL','NORMAL_SAL_IDX');

SEGMENT_NAME              Size in MB
------------------------------  ------
---------
TEST_NORMAL               50
NORMAL_SAL_IDX            17

SQL> select index_name, clustering_factor
from user_indexes;

INDEX_NAME               CLUSTERING_FACTOR
------------------------------  ------
-----------------------------
NORMAL_SAL_IDX           986778
```

Tableau 22

```
SQL> set autotrace
SQL> select * from test_normal where
sal=&sal;
Enter value for sal: 1869
old   1: select * from test_normal where
sal=&sal
new   1: select * from test_normal where
sal=1869

164 rows selected.

Elapsed: 00:00:00.01

Execution Plan
-------------------------------------------
--------------
   0      SELECT STATEMENT Optimizer=CHOOSE
(Cost=169 Card=168 Bytes=4032)
   1   0    TABLE ACCESS (BY INDEX ROWID) OF
'TEST_NORMAL' (Cost=169 Card=168 Bytes=4032)
   2   1      INDEX (RANGE SCAN) OF
'NORMAL_SAL_IDX' (NON-UNIQUE) (Cost=3
Card=168)

Statistics
-------------------------------------------
--------------
          0  recursive calls
          0  db block gets
        177  consistent gets
          0  physical reads
          0  redo size
       8461  bytes sent via SQL*Net to
client
        609  bytes received via SQL*Net from
client
         12  SQL*Net roundtrips to/from
client
          0  sorts (memory)
          0  sorts (disk)
        164  rows processed
```

Tableau 23

Dans la table ci-dessus, vous pouvez voir que cet index est supérieur à l'index bitmap sur la même colonne. Le facteur de regroupement est également proche du nombre de lignes dans cette table.

Maintenant, pour les essais; prédicats d'égalité en premier (Tableau 23), et après prédicats d'intervalles (Tableau 24).

```
SQL> select * from test_normal where sal
between &sal1 and &sal2;
Enter value for sal1: 1500
Enter value for sal2: 2000
old   1: select * from test_normal where sal
between &sal1 and &sal2
new   1: select * from test_normal where sal
between 1500 and 2000

83743 rows selected.

Elapsed: 00:00:04.03

Execution Plan
-----------------------------------------------
---------------
   0      SELECT STATEMENT Optimizer=CHOOSE
(Cost=601 Card=83376 Bytes
     =2001024)
   1    0    TABLE ACCESS (FULL) OF
'TEST NORMAL' (Cost=601 Card=83376
          Bytes=2001024)

Statistics
-----------------------------------------------
---------------
          0   recursive calls
          0   db block gets
      11778   consistent gets
       3891   physical reads
          0   redo size
    4123553   bytes sent via SQL*Net to
client
      61901   bytes received via SQL*Net from
client
       5584   SQL*Net roundtrips to/from
client
          0   sorts (memory)
          0   sorts (disk)
      83743   rows processed
```
Tableau 24

Lorsque les requêtes ont été exécutées pour des valeurs différentes, le résultat obtenu, comme indiqué dans les tableaux ci-dessous, révèle que le nombre des lectures cohérent et des lectures physiques sont identiques.

| BITMAP | | SAL (Equality) | B-TREE | | Rows Fetched |
| Consistent Reads | Physical Reads | | Consistent Reads | Physical Reads | |
|---|---|---|---|---|---|
| 165 | 0 | 1869 | 177 | 164 | |
| 169 | 163 | 3548 | 181 | 167 | |
| 174 | 166 | 6500 | 187 | 172 | |
| 75 | 69 | 7000 | 81 | 73 | |
| 177 | 163 | 2500 | 190 | 175 | |

Tableau 25

| BITMAP | | SAL (Range) | B-TREE | | Rows Fetched |
| Consistent Reads | Physical Reads | | Consistent Reads | Physical Reads | |
|---|---|---|---|---|---|
| 11778 | 5850 | 1500-2000 | 11778 | 3891 | 83743 |
| 11765 | 5468 | 2000-2500 | 11765 | 3879 | 83328 |
| 11753 | 5471 | 2500-3000 | 11753 | 3884 | 83318 |
| 17309 | 5472 | 3000-4000 | 17309 | 3892 | 166999 |
| 39398 | 5454 | 4000-7000 | 39398 | 3973 | 500520 |

Tableau 26

Pour l'intervalle de prédicats l'optimiseur opté pour un scan de table complet pour tous les différents ensemble de valeurs qu'il, n'a pas du tout utilisé les index, tandis que pour les prédicats d'égalité, l'optimiseur a utilisé les index. Encore une fois, les lectures cohérents et lectures physiques sont identiques. Par conséquent, on peut conclure pour une colonne de normale cardinal, les décisions de l'optimiseur pour les deux types d'index sont les mêmes et il n'y avait pas de différences significatives entre les E/S.

**Étape 6 (ajouter une colonne GENDER)**

Avant d'effectuer le test sur une colonne basse cardinal, nous allons ajouter une colonne de genre dans ce tableau et le mettre à jour avec M, F et les valeurs NULL.

```
SQL> alter table test_normal add GENDER
varchar2(1);

Table altered.

SQL> select GENDER, count(*) from
test_normal group by GENDER;

S      COUNT(*)
-      ---------
F      333769
M      499921
       166310

3 rows selected.
```
Tableau 27

La taille de l'index bitmap sur cette colonne est d'environ 570KB, comme indiqué dans la table ci-dessous.(Tableau 28) En revanche, l'index B-arbre sur cette colonne est 13MB en taille, ce qui est beaucoup plus grand que l'index bitmap sur cette colonne.(Tableau 29)

```
SQL> create bitmap index normal_GENDER_bmx
on test normal(GENDER);

Index created.

Elapsed: 00:00:02.08

SQL> select substr(segment name,1,30)
segment name, bytes/1024/1024 "Size in MB"
  2  from user segments
  3  where segment_name in
('TEST_NORMAL','NORMAL_GENDER_BMX');

SEGMENT NAME          Size in MB
----------------------------   -------
--------
TEST_NORMAL           50
NORMAL_GENDER_BMX     .5625

2 rows selected.
```
Tableau 28

```
SQL> create index normal_GENDER_idx on
test_normal(GENDER);

Index created.

SQL> select substr(segment_name,1,30)
segment_name, bytes/1024/1024 "Size in MB"
  2  from user segments
  3  where segment name in
('TEST NORMAL','NORMAL GENDER IDX');

SEGMENT_NAME          Size in MB
----------------------------   --------
-------
TEST NORMAL           50
NORMAL GENDER IDX     13

2 rows selected.
```
Tableau 29

Maintenant, si nous exécutons une requête avec prédicats d'égalité, l'optimiseur va pas utiliser d'index, que ce soit bitmap ou un B-arbre. Au contraire, il va préférer un scan de table complet. (Tableau 30)

```
SQL> select * from test_normal where GENDER
is null;

166310 rows selected.

Elapsed: 00:00:06.08

Execution Plan
-------------------------------------------
   0      SELECT STATEMENT Optimizer=CHOOSE
(Cost=601 Card=166310 Bytes=4157750)
   1    0    TABLE ACCESS (FULL) OF
'TEST_NORMAL' (Cost=601 Card=166310
Bytes=4157750)

SQL> select * from test normal where
GENDER='M';

499921 rows selected.

Elapsed: 00:00:16.07

Execution Plan
-------------------------------------------
   0      SELECT STATEMENT Optimizer=CHOOSE
(Cost=601 Card=499921 Bytes=12498025)
   1    0    TABLE ACCESS (FULL) OF
'TEST NORMAL' (Cost=601
Card=499921Bytes=12498025)

SQL>select * from test_normal where
GENDER='F'
 /

333769 rows selected.

Elapsed: 00:00:12.02

Execution Plan
-------------------------------------------
   0      SELECT STATEMENT Optimizer=CHOOSE
(Cost=601 Card=333769 Byte
            s=8344225)
   1    0    TABLE ACCESS (FULL) OF
'TEST NORMAL' (Cost=601 Card=333769
            Bytes=8344225)
```
Tableau 30

**V-    Conclusions :**

Maintenant que nous avons compris comment l'optimiseur réagit à ces techniques, nous allons examiner un scénario qui démontre clairement les meilleures applications respectives des index bitmap et les index B-arbres

Avec un index bitmap sur la colonne GENDER en place, on va créer un autre index bitmap sur la colonne SAL, puis d'exécuter certaines requêtes. Les requêtes seront exécutées à nouveau avec index B-arbres sur ces colonnes. Dans la table TEST_NORMAL, on aura besoin du numéro d'employé de tous les employés de sexe masculin dont les salaires mensuels égaux une des valeurs suivantes:

```
1000, 1500, 2000, 2500, 3000, 3500, 4000,
4500.
```

Ainsi :

```
SQL>select * from test_normal
where sal in
(1000,1500,2000,2500,3000,3500,4000,4500,500
0) and GENDER='M';

1453 rows selected.

Elapsed: 00:00:02.03

Execution Plan
--------------------------------------------
--------------
   0      SELECT STATEMENT Optimizer=CHOOSE
(Cost=198 Card=754 Bytes=18850)
   1    0   TABLE ACCESS (BY INDEX ROWID) OF
'TEST_NORMAL' (Cost=198 Card=754
Bytes=18850)
   2    1     BITMAP CONVERSION (TO ROWIDS)
   3    2       BITMAP AND
   4    3         BITMAP OR
   5    4           BITMAP INDEX (SINGLE
VALUE) OF 'NORMAL_SAL_BMX'
   6    4           BITMAP INDEX (SINGLE
VALUE) OF 'NORMAL_SAL_BMX'
   7    4           BITMAP INDEX (SINGLE
VALUE) OF 'NORMAL_SAL_BMX'
   8    4           BITMAP INDEX (SINGLE
VALUE) OF 'NORMAL_SAL_BMX'
   9    4           BITMAP INDEX (SINGLE
VALUE) OF 'NORMAL_SAL_BMX'
  10    4           BITMAP INDEX (SINGLE
VALUE) OF 'NORMAL_SAL_BMX'
  11    4           BITMAP INDEX (SINGLE
VALUE) OF 'NORMAL_SAL_BMX'
  12    4           BITMAP INDEX (SINGLE
VALUE) OF 'NORMAL_SAL_BMX'
  13    4           BITMAP INDEX (SINGLE
VALUE) OF 'NORMAL_SAL_BMX'
  14    3         BITMAP INDEX (SINGLE
VALUE) OF 'NORMAL_GENDER_BMX'

Statistics
--------------------------------------------
--------------
        0   recursive calls
        0   db block gets
     1353   consistent gets
      920   physical reads
        0   redo size
    75604   bytes sent via SQL*Net to
client
     1555   bytes received via SQL*Net from
client
       98   SQL*Net roundtrips to/from
client
        0   sorts (memory)
        0   sorts (disk)
     1453   rows processed
```

Tableau 31

Il s'agit d'une requête typique de l'entrepôt de données, que, bien sûr, ne faut jamais l'exécuter sur un système OLTP. Voici les résultats obtenus avec l'index bitmap en place sur deux colonnes.

Avec index B-arbre en place :

```
SQL>select * from test_normal
where sal in
(1000,1500,2000,2500,3000,3500,4000,4500,500
0) and GENDER='M';

1453 rows selected.

Elapsed: 00:00:03.01

Execution Plan
--------------------------------------------
--------------
   0      SELECT STATEMENT Optimizer=CHOOSE
(Cost=601 Card=754 Bytes=18850)
   1    0   TABLE ACCESS (FULL) OF
'TEST_NORMAL' (Cost=601 Card=754
Bytes=18850)

Statistics
--------------------------------------------
--------------
        0   recursive calls
        0   db block gets
     6333   consistent gets
     4412   physical reads
        0   redo size
    75604   bytes sent via SQL*Net to
client
     1555   bytes received via SQL*Net from
client
       98   SQL*Net roundtrips to/from
client
        0   sorts (memory)
        0   sorts (disk)
     1453   rows processed
```

Tableau 32

Bien vu, avec l'index B-tree, l'optimiseur a opté pour un scan de table complet, alors que dans le cas de l'index bitmap, il l'a utilisé pour répondre à la requête. Vous pouvez déduire les performances par le nombre d'E / S nécessaire pour aller chercher le résultat.

En résumé, les index bitmap sont les mieux adaptés pour DSS (systèmes de décisionnels) indépendamment de cardinal pour ces raisons:

- Avec index bitmap, l'optimiseur peut répondre efficacement requêtes qui incluent AND, OR, XOR ou. (Oracle supporte la conversion dynamique B-Arbre à bitmap, mais il peut être inefficace.)

- Avec bitmaps, l'optimiseur peut répondre à des questions lors de la recherche ou de comptage de zéros. Les valeurs nulles sont aussi indexées dans les index bitmap (contrairement aux index B-arbre).

- Et le plus important que les index bitmap dans les systèmes de décisionnels supportent les requêtes ad hoc, tandis que les B-arbres ne le font pas. Plus précisément, si vous avez une table avec 50 colonnes et les utilisateurs fréquemment envoient des requêtes sur 10 d'entre eux? Soit la combinaison de tous les 10 colonnes ou parfois une seule colonne? Création d'un index B-arbresera très difficile. Si vous créez 10 index bitmap sur tous ces colonnes, toutes les requêtes peuvent être traitées par ces indices, qu'ils soient des requêtes sur l'ensemble des 10 colonnes, sur 4 ou 6 colonnes sur les 10 ou sur une seule colonne. Le signe AND_EQUAL fournit cette fonctionnalité pour les index B-arbres, mais pas plus de cinq indices peut être utilisé par une requête. Cette limite n'est pas imposée avec des index bitmap.

En revanche, les index B-arbres sont bien adaptés pour les applications OLTP dans lequel les requêtes des utilisateurs sont relativement routine (et bien réglé avant le déploiement en production), par opposition à des requêtes ad hoc, qui sont beaucoup moins fréquentes et exécutés pendant heures creuses d'affaires. Comme les données sont fréquemment mis à jour et supprimés à partir d'applications OLTP, les index bitmap peuvent causer un problème de blocage sérieux dans ces situations.

Le message ici est assez clair. Les deux indices ont un objectif similaire: pour retourner des résultats aussi vite que possible. Mais le choix de celui à utiliser devrait dépendre uniquement sur le type d'application, et non sur le niveau de cardinal.

## Références

[1] S. Chaudhuri, U. Dayal,An Overview of Data Warehousing and OLAP Technology., ACM SIGMOD RECORD. 1997

[2] P. O'Neil, Model 204 Architecture and Performance. In Proceedings of the 2nd international Workshop on High Performance Transaction Systems, Lecture Notes In Computer Science, vol. 359. Springer-Verlag, London, (September 28 - 30, 1987),pp.40-59

[3] R. Kimball, L. Reeves, M. Ross, The Data Warehouse Toolkit. John Wiley Sons, NEW YORK, 2nd edition, 2002

[4] W. Inmon, Building the Data Warehouse., John Wiley Sons, fourth edition, 2005

[5] C. DELLAQUILA and E. LEFONS and F. TANGORRA, Design and Implementation of a National Data Warehouse. Proceedings of the 5th WSEAS Int. Conf. on Artificial Intelligence, Knowledge Engineering and Data Bases, Madrid, Spain, February 15-17, 2006 pp. 342-347

[6] D. Comer,Ubiquitous b-tree, ACM Comput. Surv. 11, 2, 1979, pp. 121-13

[7] R. Strohm, Oracle Database Concepts 1g,Oracle, Redwood City,CA 94065, 2007

[8] P. O'Neil and D. Quass, Improved query performance with variant indexes, In SIGMOD: Proceedings of the 1997 ACM SIGMOD international conference on Management of data.1997

[9] C. Dell aquila and E. Lefons and F. Tangorra, Analytic Use of Bitmap Indices. Proceedings of the 6th WSEAS International Conference on Artificial Intelligence, Knowledge Engineering and Data Bases, Corfu Island, Greece, February 16-19, 2007 pp. 159

[10] P. O'Neil and G. Graefe, Multi-table joins through bitmapped join indices, ACM SIGMOD Record 24 number 3, Sep 1995 , pp. 8-11.

[11] K. Wu and P. Yu, Range-based bitmap Indexing for high cardinality attributes with skew, In COMPSAC 98: Proceedings of the 22nd International Computer Software and Applications Conference. IEEE Computer Society, Washington, DC, USA, 1998, pp. 61-67.

[12] E. E-O'Neil and P. P-O'Neil, Bitmap index design choices and their performance impli-cations, Database Engineering and Applications Symposium. IDEAS 2007. 11th International, pp. 72-84.

[13] C. Imho and N. Galemmo and J. Geiger, Mastering Data Warehouse Design : Relational and Dimensional Techniques, John Wiley and Sons, NEW YORK.2003